\documentclass[
 reprint,
]{revtex4-2}

\usepackage{graphicx}
\usepackage{dcolumn,footnote}
\usepackage{amssymb,amsmath}       %
\usepackage{graphicx}                     %
\usepackage{longtable,comment}            %
\usepackage[colorlinks = true,            %
            linkcolor = blue,             %
            urlcolor  = blue,             %
            citecolor = blue,             %
            anchorcolor = blue]           %
            {hyperref}                    %

\begin{document}

\title{Random walks with homotopic spatial inhomogeneities
}

\author{Ignacio S. Gomez}
\email{
ignacio.gomez@uesb.edu.br
}
\affiliation{Departamento de Ciências Exatas e Naturais, Universidade Estadual do Sudoeste da Bahia,
			    BR 415, Itapetinga - BA, 45700-000, Brazil}

\author{Daniel Rocha de Jesus}%
 \email{202310125@uesb.edu.br}
\affiliation{Departamento de Ciências Exatas e Naturais, Universidade Estadual do Sudoeste da Bahia,
			    BR 415, Itapetinga - BA, 45700-000, Brazil}

\author{Ronaldo Thibes}%
 \email{thibes@uesb.edu.br}
\affiliation{Departamento de Ciências Exatas e Naturais, Universidade Estadual do Sudoeste da Bahia,
			    BR 415, Itapetinga - BA, 45700-000, Brazil}

\begin{abstract}
In this work we study a generalization of the standard random walk, an homotopic random walk (HRW), using a 
deformed translation unitary step that arises from a homotopy of the position-dependent masses associated to the Tsallis and Kaniadakis nonexensive statistics. The HRW implies an associated homotopic Fokker-Planck equation (HFPE) provided with a bi-parameterized inhomogeneous diffusion. 
The trajectories of the HRW exhibit convergence to a position, randomness as well as divergence, according to deformation and homotopic parameters.  
The HFPE obtained from associated master equation to the HRW presents the features: a) it results an special case of the van Kampen diffusion equation (5) of Ref. [N. G. van Kampen, \emph{Z. Phys. B Condensed Matter} \textbf{68}, 135 (1987)]; b) it exhibits a superdiffusion in function of deformation and homotopic parameters; c) Tsallis and Kaniadakis deformed FPE are recovered as special cases; d) a homotopic mixtured diffusion is observed; and e) it has a stationary entropic density, characterizing a inhomogeneous screening of the medium, obtained from a homotopic version of the H-Theorem. 
\end{abstract}

\maketitle

Random walks (RW) are mathematical models that represent several phenomena in statistical physics containing fundamental aspects of randomness, diffusion, and dynamical features of
master equations in a unified way \cite{Hughes-1995,Rudnick-2004}.
In its simplest form, the one-dimensional RW
assumes that the walker starts at $x=0$
following without restrictions (unrestricted RW) so his movement can continue within multiples of a fixed arbitrary length $l$, forward or backward, along the infinite line with no obstacles.
In turn, this shows that the RW is equivalent to a sequence of Bernoulli trials such that, if $x_i$ is the $i$th displacement and $X_i\equiv x_i/l$ is a corresponding dimensionless random variable, then the probability of $X_i=\pm 1$ is
\begin{equation}\label{RW-1}
  P(X_i=+1)=p \quad , \quad P(X_i=-1)=1-p.
\end{equation}
Here the trials are assumed independent and identically distributed as Bernoulli
variables, so we have the expectation and variance of $X_i$
\begin{equation}\label{RW-2}
  E(X_i)=2p-1 \quad , \quad \textrm{Var}(X_i)=4p(1-p)
\end{equation}
and the position after $n$ steps is
\begin{equation}\label{RW-3}
  S_n=X_1+X_2+\ldots+X_n=\sum_{i=1}^{n}X_i
\end{equation}
with $S_0=0$. 
In Fig. 1, we illustrate a possible path for the walker
after 100 steps with $x_0=0$ and $p=1/2$.
\begin{figure}[h]
\includegraphics[width=7cm]{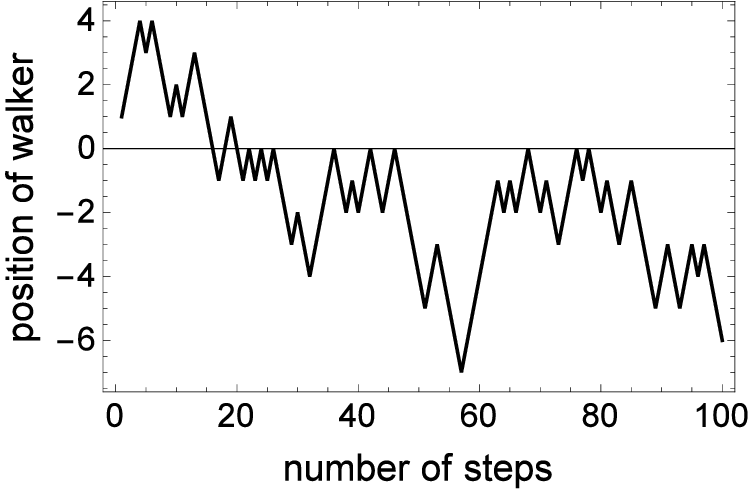}
\caption{\label{fig1}
The walker's path after $n=100$ steps, starting at $x=0$, with $p=1-p=1/2$.}
\end{figure}
Global predictions regarding the
position of the walker after $n$ steps are given by the expectation 
\begin{equation}\label{RW-4}
  E(S_n)=E\Big(\sum_{i=1}^{n}X_i\Big)=\sum_{i=1}^{n}E(X_i)=n(2p-1)
\end{equation}
and the variance corresponding to $S_n$
\begin{equation}\label{RW-5}
  \textrm{Var}(S_n)=\textrm{Var}(\sum_{i=1}^{n}X_i)
  =\sum_{i=1}^{n}\textrm{Var}(X_i)=4np(1-p).
\end{equation}

On the other hand, the progress of nonextensive statistics \cite{Tsallis-1988, Tsallis-Bukmann-1996, Baldovin-Robledo-2002, Kaniadakis-2002, Tsallis-2009, Tirnakli-Borges-2016} has driven the development of some of their associated mathematical structures, as
the $\kappa$-algebra \cite{Kaniadakis-2002} and
the $q$-algebra \cite{Nivanen-Mehaute-Wang-2003,Borges-2004}, respectively related to Kaniadakis and Tsallis statistics.
The prescriptions for the position-dependent masses (PDM) $m_{q,\kappa}(u)$
associated to $q$-algebra and $\kappa$-algebra allow to define arbitrary deformations $\eta=\eta(x)$ 
of the position $x$ given by
\cite{daCosta-PRE-2020}
\begin{equation}\label{PDM-prescription}
    \textrm{d}\eta=\sqrt{\frac{m(x)}{m_0}}\textrm{d}x.
\end{equation}
For the Tsallis PDM $m_q(x)$ case, the prescription \eqref{PDM-prescription} 
establishes a canonical trasformation map between the quantum harmonic oscillator 
with deformed momentum operator into the Morse quantum oscillator \cite{Morse-EPL}. It is worth mentioning that PDM systems have arisen originally to describe transport phenomena in semiconductor heterostructures \cite{PDM1,PDM2} and also shown importance in DFT \cite{PDM-DFT}, SUSY quantum mechanics \cite{PDM-SUSY}, nuclear physics \cite{PDM-NP}, nonlinear optics \cite{PDM-NO}, Landau quantization \cite{PDM-LQ}, superestatistical partition functions \cite{Maike-2021}, group entropy algebraic structures linked with PDM Schrödinger equations \cite{Gomez-2021}, among other fields. 

Replacing $\eta(x)$ by the $q,\kappa$-differentials of $q,\kappa$-algebra 
\begin{equation}\label{deformed-differential} 
    d_q x = \frac{dx}{1+(1-q)\xi^{-1}x} \quad , \quad 
        d_\kappa x = \frac{dx}{\sqrt{1+\kappa^2x^2}} \quad
\end{equation}
in \eqref{PDM-prescription},
we obtain
\begin{subequations}
\label{PDMs}
\begin{align}
    m_{q}(x)&=\frac{m_0}{\left[1+(1-q)\xi^{-1}x \right]^2},\\
    m_{\kappa}(x) &=\frac{m_0}{1+\kappa^2 x^2},
\end{align}
\end{subequations}
for the PDM functions respectively associated to the $q$-algebra and the $\kappa$-algebra, 
with $m_0$ standing for the constant mass case. 
In turn, from the $q,\kappa$ differentials 
\eqref{deformed-differential}
the $x_{q,\kappa}$ deformed number of $x$, the $D_{q,\kappa}$ derivatives  
and the $\int_{q,\kappa}$ integrals are defined by
\begin{subequations}\label{deformation-number}
\begin{align}
        x_q &= \int dx_q = \frac{\xi}{1-q}\ln(1+(1-q)\xi^{-1}x),\\
x_\kappa &= \int dx_\kappa = \frac{1}{\kappa}  \operatorname{arcsinh} {(\kappa x)},
\end{align}
\end{subequations}
\begin{subequations}\label{deformed-derivative}
\begin{align}
        D_q f(x)&=\frac{df}{dx_q}= (1+(1-q)\xi^{-1}x)\frac{df}{dx},\\
        D_\kappa f(x)&=\frac{df}{dx_\kappa} =\sqrt{1+\kappa^2x^2}\frac{df}{dx},
\end{align}
\end{subequations}
and 
\begin{subequations}\label{deformed-integral}
\begin{align}
        \int_q f(x)&=\int f(x)dx_q =\int f(x)\frac{dx}{1+(1-q)\xi^{-1}x},\\
        \int_\kappa f(x)&=\int f(x)dx_\kappa =\int f(x)\frac{dx}{\sqrt{1+\kappa^2x^2}},
\end{align}
\end{subequations}
Our strategy is to formulate a homotopy that connects
the PDMs (8a) and (8b) 
for introducing mixed inhomogeneities in the RW. 
We propose the homotopy
$m_{\gamma,\lambda}:X \times [0,1]\rightarrow Y$
between the PDMs corresponding to these algebraic structures
\begin{equation}
\label{eq:mlg}
m_{\gamma, \lambda}(x) = \frac{m_0}{1+2\gamma\lambda x + \gamma^2 x^2} \quad , 
\quad \lambda \in [0,1] \,,
\end{equation}
where $X,Y$ are suitable subsets of $\mathbb{R}$.
We warn that \eqref{eq:mlg} may not be strictly a homotopy 
because $X$ and $Y$ can vary according to the parameter $\gamma$.
However, the terminology is justified since 
the deformation character of the homotopy is preserved.
It should to be noted that since 
$1+2\gamma\lambda x+\gamma^2x^2=(\lambda+\gamma x)^2+1-\lambda^2$ with $1-\lambda^2\geq 0$, it follows that $1+2\gamma\lambda x+\gamma^2x^2\geq 0$ and it can be zero if and only if $\lambda=1$ and 
$x=-1/\gamma$. Physically, this says that the homotopic PDM $m_{\gamma,\lambda}(x)$ presents a divergence in $x=-1/\gamma$ only for $\lambda=1$, corresponding to the particle localization. 
From the definition \eqref{eq:mlg}, we recover the Kaniadakis and Tsallis PDMs
\begin{subequations}
\begin{align}
    m_{\gamma,0}(x) &= \frac{m_0}{1+\gamma^2 x^2}, 
		\qquad (\textrm{Kaniadakis PDM class}),\\
    m_{\gamma,1}(x) &= \frac{m_0}{(1+\gamma x)^2}, 
		\qquad (\textrm{Tsallis PDM class}).
\end{align}
\end{subequations}
where $\gamma=\kappa=(1-q)\xi^{-1}$ connects the deformation parameters of the Kaniadakis and Tsallis PDMs.
It is worth mentioning that $l$ and $\xi$ are dimensional constants playing different roles. The former represents the minimum dimensional step of the walker, while the latter is the characteristic length related to the deformation parameter $\gamma$.
Thus, the PDM prescription \eqref{PDM-prescription} applied to 
the PDM homotopy \eqref{eq:mlg} 
gives the homotopic deformation $x_{\gamma,\lambda}$ of the real number $x$
\begin{equation}
\label{eq:homotopy-deformation}
x_{\gamma,\lambda} =
\frac{1}{\gamma} \ln \left(
\frac{\lambda + \gamma x + \sqrt{1 + 2 \lambda \gamma x + \gamma^2 x^2}}{\lambda + 1} 
\right).    
\end{equation}
From \eqref{eq:homotopy-deformation}
follows the homotopic deformed 
differential 
\begin{equation}\label{homotopic-differential}
    dx_{\gamma,\lambda} = 
    \frac{dx}{\sqrt{1+2\gamma \lambda x+\gamma^2x^2}},
\end{equation}
which allows to define their corresponding 
homotopic deformed derivative 
\begin{equation}\label{homotopic-derivative}
    D_{\gamma,\lambda}=
    \frac{df}{dx_{\gamma,\lambda}}=
    \sqrt{1+2\gamma \lambda x+\gamma^2x^2}
    \frac{df}{dx},
\end{equation}
and
homotopic deformed integral 
\begin{equation}\label{homotopic-integral}
    \int_{\gamma,\lambda} f(x)
    =\int f(x)dx_{\gamma,\lambda}=
    \int f(x)\frac{dx}{\sqrt{1+2\gamma \lambda x+\gamma^2x^2}}.
\end{equation}
The $q,\kappa$-deformed differential \eqref{deformed-differential}, the $q,\kappa$-deformation of the number $x$ \eqref{deformation-number}, the $q,\kappa$-derivative \eqref{deformed-derivative} and the $q,\kappa$-integral \eqref{deformed-integral} can be promptly recovered from 
\eqref{eq:homotopy-deformation},
\eqref{homotopic-differential},
\eqref{homotopic-derivative} and 
\eqref{homotopic-integral} for the limiting cases
$\lambda=0$ and $\lambda=1$, considering respectively $\gamma=(1-q)\xi^{-1}$ and $\gamma=\kappa$.
With the aim of generating a RW with a deformed unitary step given by the homotopic deformation $x_{\gamma,\lambda}$, we recall the definition of the $G$-sum of two real numbers $a,b$ from the context of group entropies and their mathematical associated structures \cite{Gomez-2021,Tempesta-2011}
\begin{equation}\label{G-property}
    a \oplus_G b =G\Big(G^{-1}(a)+G^{-1}(b)\Big) \,,
\end{equation}
where $G^{-1}(a)=a_G$ is the $G$-deformation of $a$ corresponding to the $G$-group class. We employ the homotopic deformation \eqref{eq:homotopy-deformation}  characterizing $x_G$ in \eqref{G-property}, that is 
\begin{equation}\label{homotopic-sum}
    a \oplus_{\gamma,\lambda} b =\Big(a_{\gamma,\lambda}+b_{\gamma,\lambda}\Big)^{-1}\,,
\end{equation}
with
the corresponding inverse transformation 
\begin{equation}
\label{eq:inverse-homotopy-deformation}
x_{\gamma,\lambda}^{-1} =
\frac{\lambda[\cosh (\gamma x_{\gamma,\lambda} )-1]
	+\sinh (\gamma{x_{\gamma,\lambda}})}{\gamma}\,.    
\end{equation}
Using \eqref{eq:homotopy-deformation} and \eqref{eq:inverse-homotopy-deformation}, we can write the homotopic sum for the present case explicitly as 
\begin{equation}\label{homotopic-sum-2}
    x \oplus_{\gamma,\lambda} y = \frac{\Big(\frac{\lambda+1}{2}\Big)e^{\gamma(x_{\gamma,\lambda}+y_{\gamma,\lambda})}+\Big(\frac{\lambda-1}{2}\Big)e^{-\gamma(x_{\gamma,\lambda}+y_{\gamma,\lambda})}-\lambda}{\gamma} \,.
\end{equation}

In the same way as in the recent work \cite{GomezPRE-2023}, for preserving consistency, we postulate the probability of the homotopic
deformed position $(x_i)_{\gamma,\lambda}$ as
being the same of the RW in standard space \eqref{RW-1}, that is
\begin{equation}\label{homotopic-probability}
  P((x_i)_{\gamma,\lambda}=l_{\gamma,\lambda})=p \ \ , \ \ P((x_i)_{\gamma,\lambda}=(-l)_{\gamma,\lambda})=1-p.
\end{equation}
with the homotopic deformed unitary steps resulting from \eqref{eq:homotopy-deformation}
\begin{subequations}\label{homotopic-steps}
\begin{align}
l_{\gamma,\lambda} =
\frac{1}{\gamma} \ln \left(
\frac{\lambda + \gamma l  + \sqrt{1 + 2 \lambda \gamma l + \gamma^2 l^2 }}{\lambda + 1} 
\right)\,, \\
    (-l)_{\gamma,\lambda} =
\frac{1}{\gamma} \ln \left(
\frac{\lambda - \gamma l  + \sqrt{1 - 2 \lambda \gamma l  + \gamma^2 l^2 }}{\lambda + 1} 
\right)\,.
  \end{align}
\end{subequations}
Equations (23a) and (23b) mean that
we are assuming a RW dynamics in the homotopic deformed space $x_{\gamma,\lambda}$.

We can deform homotopically the 
position after $n$ steps $s_n$ by applying repeatedly \eqref{homotopic-sum}, so we obtain
\begin{eqnarray}\label{homotopic-trajectory}
    s_{n,\gamma,\lambda}=
    x_1 \oplus_{\gamma,\lambda} \ldots \oplus_{\gamma,\lambda}
    x_n =  \oplus_{\gamma,\lambda}x_i = \nonumber\\ =\frac{\Big(\frac{\lambda+1}{2}\Big)e^{\gamma\sum_{i=1}^n(x_i)_{\gamma,\lambda}}+\Big(\frac{\lambda-1}{2}\Big)e^{-\gamma\sum_{i=1}^n(x_i)_{\gamma,\lambda}}-\lambda}{\gamma}=\nonumber\\
    =\frac{\Big(\frac{\lambda+1}{2}\Big)e^{\gamma nE((x_i)_{\gamma,\lambda})}+\Big(\frac{\lambda-1}{2}\Big)e^{-\gamma nE((x_i)_{\gamma,\lambda})}-\lambda}{\gamma}, \nonumber\\   
\end{eqnarray}
where $E((x_i)_{\gamma,\lambda})$ is the expectation of $(x_i)_{\gamma,\lambda}$ given by 
\begin{eqnarray}
\label{homotopic-expectation}
\frac{1}{n}\sum_{i=1}^n(x_i)_{\gamma,\lambda}\longrightarrow
E((x_i)_{\gamma,\lambda})=p l_{\gamma,\lambda}+(1-p)(-l)_{\gamma,\lambda} \nonumber\\
\end{eqnarray}
that naturally depends on the probability $0\leq p\leq 1$, the deformation parameter $\gamma$ and the homotopic parameter $\lambda$.  To avoid divergences, from now on we consider only values for the deformation parameter $\gamma$ restricted to the physical condition $|\gamma|<\xi$.
We can establish the following result regarding the HRW  homotopic path $s_{n,\gamma,\lambda}$.

\noindent \emph{\underline{Theorem} (characterization of the HRW trajectories): The homotopic trajectories $s_{n,\gamma,\lambda}$ given by \eqref{homotopic-trajectory} satisfy:
\begin{itemize}
    \item[$(I)$] The asymptotic behavior of $s_{n,\gamma,\lambda}$ is determined by 
\begin{equation}\label{homotopic-trajectory-formula}
        s_{n,\gamma,\lambda} \longrightarrow \frac{\Big(\frac{\lambda+1}{2}\Big)e^{-n\tau_{\gamma,\lambda}}+\Big(\frac{\lambda-1}{2}\Big)e^{n\tau_{\gamma,\lambda}}-\lambda}{\gamma} 
    \end{equation}
  for $n\gg1$, with $\tau_{\gamma,\lambda}$ a (dimensionless) homotopic characteristic time defined as
    \begin{eqnarray}
     \tau_{\gamma,\lambda} &=&
    -\ln\Bigg[\Bigg(\frac{\lambda + \gamma l  + \sqrt{1 + 2 \lambda \gamma l  + \gamma^2 l^2 }}{\lambda + 1}\Bigg)^p
    \nonumber\\&&
    \Bigg(\frac{\lambda - \gamma l  + \sqrt{1 - 2 \lambda \gamma l  + \gamma^2 l^2 }}{\lambda + 1}\Bigg)^{1-p}\Bigg] \label{homotopic-time}
    \end{eqnarray}
    \item[$(II)$] For $\gamma\rightarrow 0$ we recover the standard trajectory of the RW, that is 
    \begin{eqnarray}
        s_{n,\gamma,\lambda} \longrightarrow \sum_{i=1}^{n}x_i
    \end{eqnarray}
    \item[$(III)$] For an equilibrated HRW ($p=1/2$) $\tau_{\gamma,\lambda}$ is invariant against the change of sign of the deformation $\gamma
    \rightarrow -\gamma$. 
    \item[$(IV)$] For $\lambda=1$ (Tsallis PDM) we have \begin{eqnarray}\label{trajetory-Tsallis}
    s_{n,\gamma,1}\longrightarrow\frac{e^{-n\tau_{\gamma,1}}-1}{\gamma}\quad , \quad \textrm{for $n\gg1$} \,,
    \end{eqnarray}
    with
\begin{eqnarray}\label{time-Tsallis}
        \tau_{\gamma,1}=-\ln\Bigg(\Bigg(\frac{1+\gamma l +|1+\gamma l|}{2}\Bigg)^p\Bigg(\frac{1-\gamma l +|1-\gamma l|}{2}\Bigg)^{1-p}\Bigg) \quad . \nonumber\\
    \end{eqnarray}
    For an equilibrated HRW ($p=1/2$) we have that, if $|\gamma|<l^{-1}$ then
    $\tau_{\gamma,1}=-(1/2)\ln(1-\gamma^2 l^2)>0$. In turn, this implies that
        $s_{n,\gamma,1}$ converges to $-1/\gamma$.
    \item[$(V)$] For $\lambda=0$ (Kaniadakis PDM) we have \begin{eqnarray}\label{trajetory-Kaniadakis}
    s_{n,\gamma,0}\longrightarrow\frac{\textrm{sinh}(-n\tau_{\gamma,0})}{\gamma}\quad , \quad \textrm{for $n\gg1$} 
    \end{eqnarray}
    with 
    \begin{eqnarray}\label{time-Kaniadakis}
        \tau_{\gamma,0}=-\ln\Bigg(\Bigg(\gamma l +\sqrt{1+\gamma^2 l^2}\Bigg)^p\Bigg(-\gamma l +\sqrt{1+\gamma^2 l^2 }\Bigg)^{1-p}\Bigg) . \nonumber
        \\
    \end{eqnarray}
    For an equilibrated HRW ($p=1/2$) we have 
    $\tau_{\gamma,0}=0$. In turn, this implies that
        $s_{n,\gamma,0}$ converges to zero. Thus, randomness is preserved. 
    \item[$(VI)$] For $\lambda\neq 1$, $|\gamma|<l^{-1}$ and $p=1/2$, the homotopic characteristic time $\tau_{\gamma,\lambda}\in \mathbb{R}$, so $\lim_{n\rightarrow\infty}s_{n,\gamma,\lambda}=\infty$. That is, all trajectory diverges when $n\rightarrow \infty$.  
\end{itemize}
}
\noindent\underline{Proof}:
(I): Inserting explicitly the expressions of Eqns. (23a) and (23b) into Eq. \eqref{homotopic-expectation} and then substituting the emerging expectation value $E((x_i)_{\gamma,\lambda})$ into Eq. \eqref{homotopic-trajectory},  the result follows immediately, with the homotopic characteristic time defined by $\tau_{\gamma,\lambda}=-\gamma E((x_i)_{\gamma,\lambda})$.   \\
(II): It is sufficient to see that the homotopic sum satisfies $x\oplus_{\gamma,\lambda}y\rightarrow x+y$ when $\gamma\rightarrow 0$.  Indeed, from Eq. \eqref{eq:homotopy-deformation} taking the limit $\gamma\rightarrow 0$, we have 
\begin{eqnarray}\label{proof-II1}
    \lim_{\gamma\rightarrow0}x_{\gamma,\lambda}=\lim_{\gamma\rightarrow0} \frac{1}{\gamma} \ln  \left(
\frac{\lambda + \gamma x + \sqrt{1 + 2 \lambda \gamma x + \gamma^2 x^2}}{\lambda + 1} 
\right) = \nonumber\\
\lim_{\gamma\rightarrow0} \frac{1}{\gamma}
\ln  \left(
\frac{\lambda + 1 + (1+\lambda)\gamma x}{\lambda + 1} 
 \right)=
\lim_{\gamma\rightarrow0} \frac{1}{\gamma}
\ln  \left(
{ 1+\gamma x} 
 \right)=x \nonumber
\end{eqnarray}
and then 
\begin{eqnarray}\label{proof-II2}
\lim_{\gamma\rightarrow0} 
x \oplus_{\gamma,\lambda} y=
\lim_{\gamma\rightarrow0}
\frac{\Big(\frac{\lambda+1}{2}\Big)e^{\gamma(x_{\gamma,\lambda}+y_{\gamma,\lambda})}}{\gamma} + \nonumber  \\
+\lim_{\gamma\rightarrow0} \frac{\Big(\frac{\lambda-1}{2}\Big)e^{-\gamma(x_{\gamma,\lambda}+y_{\gamma,\lambda})}-\lambda}{\gamma} = \nonumber \\
\lim_{\gamma\rightarrow0}
\frac{\gamma (x_{\gamma,\lambda}+y_{\gamma,\lambda})}{\gamma}=
\lim_{\gamma\rightarrow0}
(x_{\gamma,\lambda}+y_{\gamma,\lambda})=x+y
\end{eqnarray}
which completes the proof of (II). 
Additionally, we can also show that, for $\gamma l<<1$, we have
\begin{equation}
 \tau_{\gamma,\lambda} \rightarrow  -p \ln [1+\gamma l] - (1-p) \ln[1-\gamma l] \rightarrow (1-2p)\gamma l \,.
\end{equation}
Hence, for fixed $l$
\begin{equation}\nonumber
\lim_{\gamma \rightarrow 0} s_{n,\gamma,\lambda}= 
\end{equation}
\begin{eqnarray}\nonumber
= \lim_{\gamma  \rightarrow 0} \frac{(\lambda+1)(1-n[(1-2p)\gamma l])}{2\gamma} + \\
+\lim_{\gamma  \rightarrow 0} \frac{
(\lambda-1)(1+n[(1-2p)\gamma l]}{2\gamma} \nonumber
\end{eqnarray}
\begin{equation}
=  n (2p-1) l \,,
\end{equation}
agreeing with \eqref{RW-4}.  
\\
(III): It is sufficient to see that, for $p=1/2$, the product in the logarithm argument in \eqref{homotopic-time} is clearly invariant against the transformation $\gamma\rightarrow-\gamma$. \\
(IV): Substituting $\lambda=1$ in \eqref{homotopic-trajectory-formula},
we obtain \eqref{trajetory-Tsallis}. Using the fact $\sqrt{1\pm2\gamma l +\gamma^2 l^2 }=|1\pm \gamma l|$ with $\lambda=1$ in \eqref{homotopic-time},  expression \eqref{time-Tsallis}  follows  for $\tau_{\gamma,1}$. Also, if $|\gamma|<l^{-1}$ then $|1\pm \gamma l|=1\pm \gamma l$ so we have 
$(1\pm \gamma +|1\pm \gamma l|)/2=1\pm \gamma l$. Thus, letting $p=1/2$ in \eqref{time-Tsallis} then we obtain $\tau_{\gamma,1}=-(1/2)\ln(1-\gamma^2 l^2)>0$. From \eqref{trajetory-Tsallis}, this shows the convergence of $s_{n,\gamma,1}$ to $-1/\gamma$. \\
(V): Substituting $\lambda=0$ in \eqref{homotopic-trajectory-formula},
we obtain \eqref{trajetory-Kaniadakis}. Also, replacing $\lambda=0$ in \eqref{homotopic-time} leads to \eqref{time-Kaniadakis}.
In turn, considering $p=1/2$ in 
\eqref{time-Kaniadakis},
we obtain $\tau_{\gamma,0}=-(1/2)\ln((\gamma l +\sqrt{1+\gamma^2 l^2})(-\gamma l +\sqrt{1+\gamma^2 l^2}))=-(1/2)\ln(-\gamma^2 l^2+1+\gamma^2 l^2)=0$. From \eqref{trajetory-Kaniadakis} this shows the convergence of $s_{n,\gamma,0}$ to zero as the standard RW case.
\\
(VI): If $\lambda \neq 1$ then $0<\lambda<1$ so the two terms in \eqref{homotopic-trajectory-formula} proportional to $e^{-n\tau_{\gamma,\lambda}}$ and $e^{n\tau_{\gamma,\lambda}}$ survive in the asymptotic limit of $s_{n,\gamma,\lambda}$. Since $\tau_{\gamma,\lambda}$ is real and finite for $\lambda\in [0,1]$, $|\gamma|<l^{-1}$ and $p=1/2$ it follows that $e^{-n\tau_{\gamma,\lambda}}\longrightarrow \infty$ and  $e^{n\tau_{\gamma,\lambda}}\longrightarrow 0$ or vice-versa. Therefore, $s_{n,\gamma,\lambda}\longrightarrow \infty$ and the trajectory diverges. $\square$

\begin{figure}[h]
\centering
\includegraphics[width=1\linewidth]{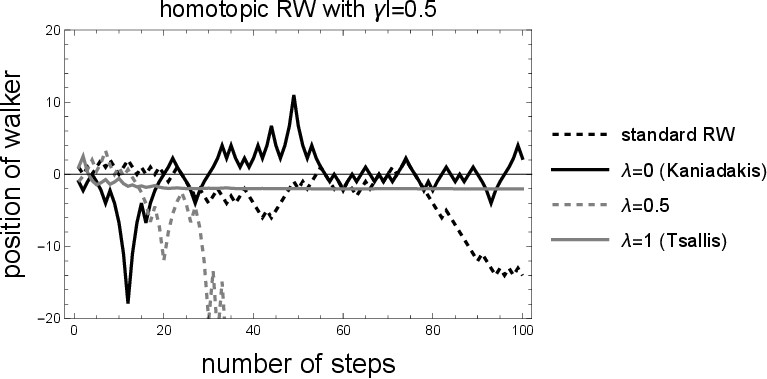} \\
\vspace{0.2cm}
\includegraphics[width=1\linewidth]{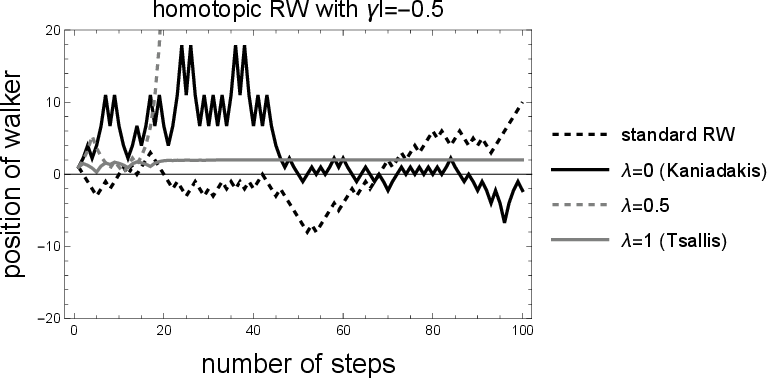}
\caption{\label{fig2}
Paths of the homotopic walker after $n=100$ steps starting at $x=0$ with $p=1/2$ for the standard (dashed black line), Kaniadakis (solid black line), mixed (dashed gray line) and Tsallis (solid gray line) homotopic classes of the random walk. Convergence to the position $x=-1/\gamma$ is reported only for the Tsallis class, in agreement with Theorem (IV). According to Theorem (V), we see that randomness is preserved for Kaniadakis class as in the standard case.
}
\end{figure}

In Fig. 2, we show some trajectories for the HRW and the standard RW for $|\gamma l|=0.5$, for the Kaniadakis, mixture and Tsallis classes.
Randomness is reported for the standard case and Kaniadakis class, while a convergence to $x=-1/\gamma$ is only observed for the Tsallis class, in agreement with Theorem (IV) and (V).

Complementarily, the connection between discrete microscopic dynamics of the HRW and its counterpart diffusion dynamics is accomplished by noticing that the probability of the particle to be in a certain position is independent of the frame used. 
If $P(x,t)$ and $\mathcal{P}(x_{\gamma,\lambda},t)$ represent the probability density in standard space $x$ and in homotopic deformed space $x_{\gamma,\lambda}$, this physical assumption is expressed by the relation  
\begin{eqnarray}\label{normalization}
    P(x,t)dx = \mathcal{P}(x_{\gamma,\lambda},t)dx_{\gamma,\lambda} \nonumber\\ 
    \Longrightarrow P(x,t)=\frac{\mathcal{P}(x_{\gamma,\lambda}(x),t)}{\sqrt{1+2\gamma \lambda x+\gamma^2x^2}},
\end{eqnarray}
where we have used \eqref{homotopic-differential} for $dx_{\gamma,\lambda}$. 
Moreover, equation  
\eqref{normalization} establishes that the normalization of $P(x,t)$ in standard space implies the normalization of $\mathcal{P}(x_{\gamma,\lambda},t)$ in homotopic deformed space $x_{\gamma,\lambda}$ and vice-versa. 

The next step is to obtain the diffusion equation for the HRW, whose motivations lie in a growing interest in inhomogeneous Fokker Planck equations due their versatility for characterizing lattice models \cite{Wu-2012}, diffusion in time-space fractional contexts \cite{Gomez-Aguilar-2016}, Lévy processes \cite{Pinto-2017}, non-singular kernel operators \cite{Maike-2018} etc.  
We consider the master equation for a standard random walk in a one-dimensional lattice expressed in the homotopic deformed space $x_{\gamma,\lambda}$, with $(+l)_{\gamma,\lambda}$ and $(-l)_{\gamma,\lambda}$ the two possible jumps that can occur in regular
intervals of time $\Delta t$, with $p$ and $1-p$ the probabilities for a right $(+l)_{\gamma,\lambda}$ or
left $(-l)_{\gamma,\lambda}$ jumps, with $l$ the parameter of the lattice in standard space $x$ {(or equivalently, the minimal length of the walker)}. 
The master equation for $\mathcal{P}(x_{\gamma,\lambda},t)$ 
is obtained by a similar method as performed in \cite{Renio-2017,Jose-2022}, but expressed in the $x_{\gamma,\lambda}$ space, which generalizes the one employed in $x_\kappa$ space \cite{CNSNS-2023}, 
\begin{eqnarray}\label{DME}
  &\mathcal{P}(x_{\gamma,\lambda},t+\Delta t)=\\
  &p\mathcal{P}(x_{\gamma,\lambda}+(+l)_{\gamma,\lambda},t)+(1-p)\mathcal{P}(x_{\gamma,\lambda}+(-l)_{\gamma,\lambda},t). \nonumber
\end{eqnarray}
The passage to the continuum in (\ref{DME}) can be accomplished by taking the limit $l\rightarrow0$.
For the sake of simplicity, we consider the equilibrated case $p=1/2$ in
(\ref{DME}). When $l\rightarrow0$, we have the following approximations up to second order terms
\begin{eqnarray}\label{approx}
  (\pm l)_{\gamma,\lambda}
\approx \pm l \nonumber\\
  \Delta \mathcal{P}_+\approx (+l)_{\gamma,\lambda} D_{\gamma,\lambda}\mathcal{P}+
  \frac{1}{2} ((+l)_{\gamma,\lambda})^2 D_{\gamma,\lambda}^2\mathcal{P}\nonumber\\
  \Delta \mathcal{P}_{-}\approx (-l)_{\gamma,\lambda} D_{\gamma,\lambda}\mathcal{P}+
  \frac{1}{2} ((-l)_{\gamma,\lambda})^2 D_{\gamma,\lambda}^2\mathcal{P}\nonumber \\
  \Delta \mathcal{P}_t \approx
  \Delta t \frac{\partial \mathcal{P}}{\partial t},
\end{eqnarray}
with $\Delta \mathcal{P}_+=\mathcal{P}(x_{\gamma,\lambda}+(+l)_{\gamma,\lambda},t)-\mathcal{P}(x_{\gamma,\lambda},t)$, $\Delta \mathcal{P}_{-}=\mathcal{P}(x_{\gamma,\lambda}+(-l)_{\gamma,\lambda},t)-\mathcal{P}(x_{\gamma,\lambda},t)$ and 
$\Delta \mathcal{P}_t=\mathcal{P}(x_{\gamma,\lambda},t+\Delta t)-\mathcal{P}(x_{\gamma,\lambda},t)$.
Substituting (\ref{approx})
in (\ref{DME}), with $p=1/2$, we arrive at the Fokker Planck equation \cite{Oppenheim-1977,Risken-1989} in the context of a homotopy between the Tsallis and Kaniadakis PDMs (that we called briefly HFPE)
with a null drift term
\begin{equation}\label{HFPE}
  \frac{\partial \mathcal{P}(x_{\gamma,\lambda},t)}{\partial t}=\Gamma D_{\gamma,\lambda}^2\mathcal{P}(x_{\gamma,\lambda},t),
\end{equation}
with $\Gamma=l^2/(2\Delta t)$ the diffusion coefficient of the
standard case.
The HFPE written in standard space
\begin{equation}
\label{HFPE-standard}
  \frac{\partial P(x,t)}{\partial t}=
  \Gamma \frac{\partial}{\partial x}
  \sqrt{1+2\gamma\lambda x+\gamma^2x^2}\frac{\partial}{\partial x}
  \sqrt{1+2\gamma\lambda x+\gamma^2x^2}P(x,t)
\end{equation}
results in a particular case of the van Kampen diffusion equation (5) of \cite{vanKampen-1987}
\begin{equation}\label{van-Kampen}
  \frac{\partial P(x,t)}{\partial t}=
  \frac{\partial}{\partial x}[\mu(x)V'(x)P(x,t)]
  +
  \frac{\partial}{\partial x}
  \mu(x)\frac{\partial}{\partial x}
  T(x)P(x,t)
  \,,
\end{equation}
with $V'(x)=0$ and the conditions
\begin{equation}\label{conditions}
  \frac{\mu(x)}{\mu_0}=\frac{T(x)}{T_0}=\sqrt{1+2\gamma\lambda x+\gamma^2x^2} \quad , \quad \Gamma=\mu_0T_0\,,
\end{equation}
for mobility and temperature featured as position-dependent, with
$\mu_0,T_0$ denoting the values corresponding to
the homogeneous case.

It should be noted that
the homotopic solution $\mathcal{P}(x_{\gamma,\lambda},t)$ of the HFPE \eqref{HFPE}
corresponds to the free particle in the homotopic deformed space $x_{\gamma,\lambda}$
\begin{equation}\label{free-HFPE-1}
  \mathcal{P}(x_{\gamma,\lambda},t)=\frac{1}{\sqrt{2\pi \Gamma t}}\exp\left(-\frac{x_{\gamma,\lambda}^2}{2\Gamma t} \right)
\end{equation}
then using (\ref{normalization}), it follows the solution $P(x,t)$
\begin{eqnarray}\label{free-HFPE-2}
  P(x,t)= &\\
  \frac{1}{\sqrt{1+2\gamma\lambda x+\gamma^2x^2}} 
  \frac{1}{\sqrt{2\pi \Gamma t}} & \exp\left(-\frac{\ln^2 \left(
\frac{\lambda + \gamma x + \sqrt{1 + 2 \lambda \gamma x + \gamma^2 x^2}}{\lambda + 1} 
\right)}{(2\Gamma t)\gamma^2} \right). \nonumber 
\end{eqnarray}
In order to proceed our analysis, from now on we set $\Gamma T/\xi^2=1$ and $\gamma \xi=\pm0.5$, {where $T$ is a constant with time dimension}. In Fig. 3, we illustrate $P(x,t)$ for times
$t/T=0.5, 2$
and for $\gamma \xi=\pm0.5$ with the initial condition
$P(x,t=0)=\delta(x)$. We can see that the homotopy introduces a mixture of probability density distributions between the Kaniadakis ($\lambda=0$) and Tsallis ($\lambda=1$) cases, removing the Tsallis divergence at $x=-1/\gamma$ for any value of $\lambda$ such that $0<\lambda<1$. Asymmetry and divergence is observed only for the Tsallis case as a consequence of its PDM functional form (\ref{PDMs}a). The Tsallis divergence at $x=-1/\gamma$ is cancelled as a result of the homotopy for $0<\lambda<1$.   
\begin{figure}[h]
\centering
\includegraphics[width=1\linewidth]{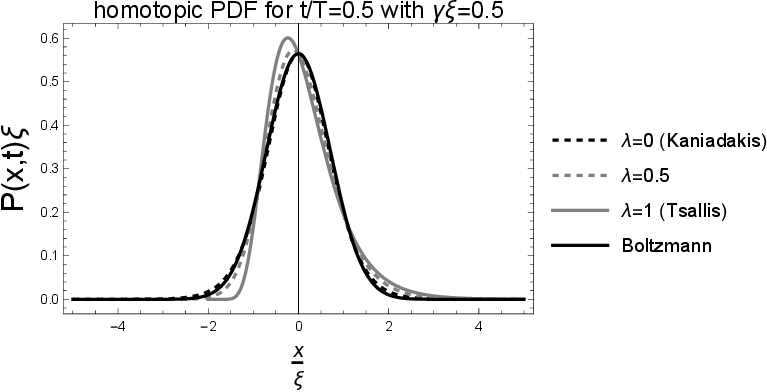} \\
\vspace{0.2cm}
\includegraphics[width=1\linewidth]{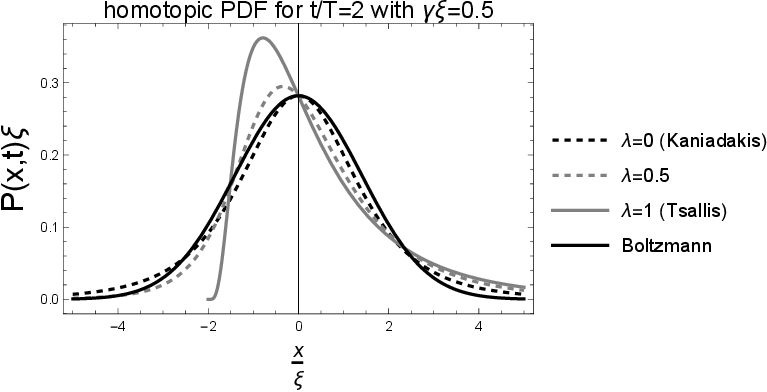}
\caption{\label{fig3}
Evolution of the dimensionless homotopic probability density distribution given by \eqref{free-HFPE-2} for the Kaniadakis, mixture, Tsallis and standard (Boltzmann) cases with delta initial condition $P(x,t=0)=\delta(x)$ and $\gamma \xi=0.5$. }
\end{figure}

\begin{figure}[h]
\centering
\includegraphics[width=1\linewidth]{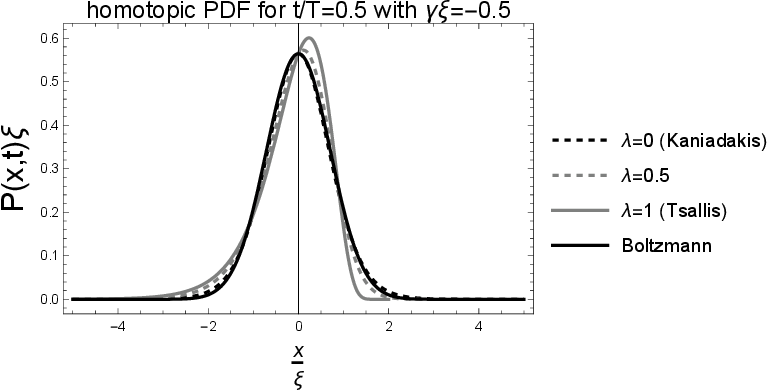} \\
\vspace{0.2cm}
\includegraphics[width=1\linewidth]{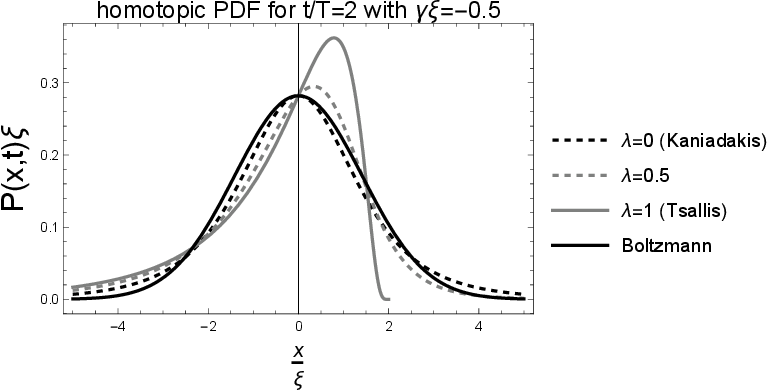}
\caption{\label{fig4}
Evolution of the dimensionless homotopic probability density distribution given by \eqref{free-HFPE-2} for the Kaniadakis, mixture, Tsallis and standard (Boltzmann) cases with delta initial condition $P(x,t=0)=\delta(x)$ and $\gamma \xi=-0.5$.
}
\end{figure}

Next, to investigate further the diffusional aspects 
of the HFPE,  we calculate the mean standard deviation (MSD) for the solution $P(x,t)$ given by \eqref{free-HFPE-2}: 
\begin{eqnarray}\label{HMSD}
\sigma_{\gamma,\lambda}^2= \langle x^2(t) \rangle - \langle x(t) \rangle^2=\nonumber\\
=\int_{-\infty}^{\infty}x^2 P(x,t)dx -
    \Bigg(\int_{-\infty}^{\infty}x P(x,t)dx\Bigg)^2,
\end{eqnarray}
a quantity which we name as the homotopic mean standard deviation (HMSD). In Fig. 5, we plot numerically the HMSD for $t/T\in [0,5]$. For $0<\lambda<1$, a superdiffusion behavior (above normal diffusion) is observed, which manifests a fast spreading of the particle as a consequence of the homotopy between the inhomogeneous diffusion-types of Kaniadakis and Tsallis cases. HSMD is symmetrical against the change of sign $\gamma \rightarrow -\gamma $.

\begin{figure}[h]
\centering
\includegraphics[width=1\linewidth]{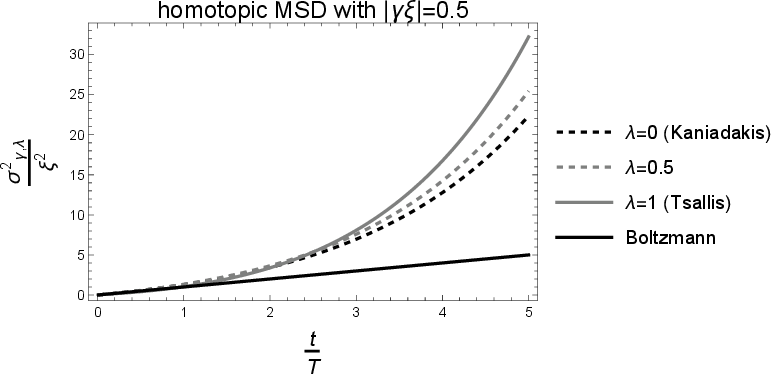}
\caption{\label{fig5}
Evolution of the dimensionaless HMSD given by \eqref{free-HFPE-2} and \eqref{HMSD} for the Kaniadakis, mixture, Tsallis and standard (Boltzmann) cases with delta initial condition $P(x,t=0)=\delta(x)$ and $|\gamma \xi|=0.5$. Mixture case between Kaniadakis and Tsallis classes indicates the presence of the homotopy. 
}
\end{figure}

For the purpose of obtaining the associated entropy functional $\mathcal{S}$
for $P(x,t)$, we notice that the HFPE \eqref{HFPE} satisfies the
$H$-Theorem in the homotopic deformed frame for $\mathcal{S}$ given by the standard Boltzmann-Gibbs form but expressed in the  homotopic coordinate space $x_{\gamma,\lambda}$. That is, 
\begin{equation}
\label{homotopic-entropy}
 \mathcal{S}
 = - \int
    \mathcal{P}(x_{\gamma,\lambda},t)
    \ln \mathcal{P} (x_{\gamma,\lambda},t) dx_{\gamma,\lambda}.
\end{equation}
In turn, by expressing $\mathcal{S}$ in the standard space $x$ we obtain \begin{eqnarray}
 \label{homotopic-entropy-2}
 \mathcal{S} &=& -\int P(x,t) \ln \left[\sqrt{1+2\gamma\lambda x+\gamma^2x^2}P(x,t)\right] dx
 \\
             &=& S_{\text{BG}} - \left\langle \ln \sqrt{1+2\gamma\lambda x+\gamma^2x^2} \right\rangle=\int s[P(x,t)]dx, \nonumber
\end{eqnarray}
with $s[P(x,t)]$ the entropic density belonging to $\mathcal{S}$.
The homotopic entropy $\mathcal{S}$ 
is composed by the sum of two terms: the Boltzmann-Gibbs entropy
$S_{\text{BG}} = -\int P(x,t) \ln P(x,t) dx$
characterizing the distribution of the particles plus a contribution representing the homotopic inhomogeneities of the PDM Tsallis and Kaniadakis classes
$S_{\text{medium}} = -\int P(x,t) \ln \sqrt{1+2\gamma\lambda x+\gamma^2x^2}dx $.
The quantity $\mathcal{S}$ resembles 
the Kullback-Leibler divergence,
or relative entropy,
$S_{\text{KL}}(P,P_0)= -\int P(x,t)\ln\left[P(x, t)/P_0(x,t)\right]dx$,
with the reference distribution
$P_0(x,t)$ replaced by 
the function
$1/\sqrt{1+2\gamma\lambda x+\gamma^2x^2}$. 
For the 
asymptotic limit
of the free particle case $P_{\textrm{stat}}(x)=\lim_{t\rightarrow\infty}P(x,t)$, by employing that $\lim_{t\rightarrow\infty}\mathcal{P}(x_{\gamma,\lambda},t)=K$ with $\mathcal{P}(x_{\gamma,\lambda},t)=P(x,t)\sqrt{1+2\gamma\lambda x+\gamma^2x^2}$, we have the stationary solution $P_{\textrm{stat}}(x)=K/\sqrt{1+2\gamma\lambda x+\gamma^2x^2}$. By replacing it in
\eqref{homotopic-entropy-2} we obtain the homotopic entropic density in the asymptotic limit
\begin{eqnarray}\label{homotopic-entropic-density-stationary}
  s[P_{\textrm{stat}}]=-K\ln K
  -K\ln\sqrt{1+2\gamma\lambda x+\gamma^2x^2}\nonumber\\
  =s_{BG}-K\ln\sqrt{1+2\gamma\lambda x+\gamma^2x^2},
\end{eqnarray}
with the second term representing the effect of the inhomogeneous medium modelled by the homotopic deformation and $s_{BG}=-K\ln K$ the Boltzmann-Gibbs entropic density. In Fig. 6, we illustrate the 
stationary entropic density (SED)
ratio $s[P_{\textrm{stat}}]/s_{BG}$ 
for the free particle case with
the deformation parameter $\gamma \xi=\pm0.5$, $x/\xi\in [-10,10]$ and for the Kaniadakis, mixture, Tsallis and Boltzmann cases. For the sake of simplicity, we have taken $\ln K=1$. 
Interesting features of the SED are observed: the asymmetry for $0<\lambda<1$, the divergence at $x=-1/\gamma$ for the Tsallis case along with its cut-off for $x\geq -1/\gamma$, the symmetry of Kaniadakis case around $x=0$ and the mixture case placed between Kaniadakis and Tsallis classes. Homotopic inhomogeneities of the medium persist in the asymptotic limit $t\rightarrow \infty$ expressed by the dependence of the SED with the position $x$ of the free particle, thus behaving like a screening effect. 

\begin{figure}[h]
\centering
\includegraphics[width=1\linewidth]{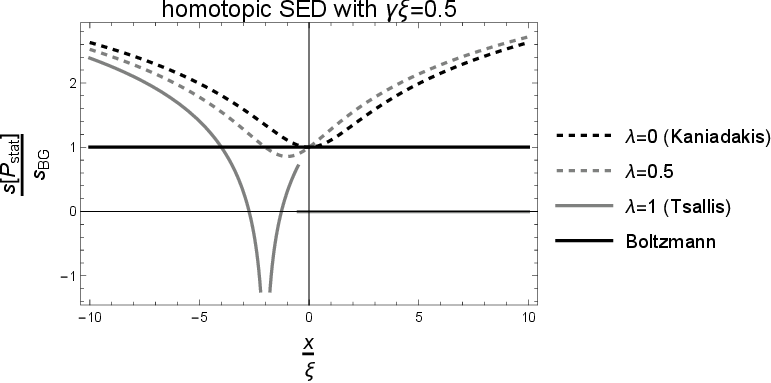} \\
\vspace{0.2cm}
\includegraphics[width=1\linewidth]{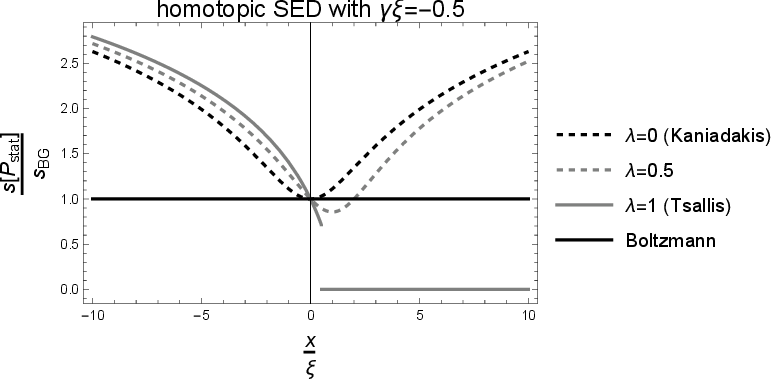}
\caption{\label{fig6}
Homotopic stationary entropic density ratio \eqref{homotopic-entropic-density-stationary} for the Kaniadakis, mixture, Tsallis and standard (Boltzmann) cases with delta initial condition $P(x,t=0)=\delta(x)$ and $|\gamma \xi|=0.5$. The change of sign $\gamma \rightarrow -\gamma $ implies a displacement of the homotopic SED from the left to the right side of $x=0$ as a consequence of the asymmetric Tsallis PDM.   
}
\end{figure}

We have presented a generalization for the random walk, named here homotopic random walk (HRW), by employing a deformed translation of the unitary step that arises by linking homotopically the position-dependent masses of the $q$-algebra and the $\kappa$-algebra, inherited by the Tsallis and Kaniadakis nonextensive statistics. Complementarily, we have obtained the associated homotopic Fokker Planck equation (HFPE) to the HRW, which expresses the homotopic inhomogeneous microscopic features of the HRW in the continuous limit. In the following, we enumerate the contributions of our work.
\begin{itemize}

\item[$(a)$]  We have obtained a theoretical characterization for the trajectories of the HRW by means of Theorem (I)--(VI), which gives account of analytical asymptotic expressions for the homotopic position $s_{n,\gamma,\lambda}$ and characteristic time $\tau_{n,\gamma,\lambda}$  after $n$ steps (with $n\gg1$), as a simple consequence of the $G$-sum property \eqref{homotopic-sum}.

\item[$(b)$] Trajectories of the HWR present a convergence to a position $x=-1/\gamma$ with $\gamma \in \mathbb{R}-\{0\}$ for Tsallis class ($\lambda=1$) and randomness behavior for Kaniadakis class ($\lambda=0$) with $\gamma \in \mathbb{R}-\{0\}$, according to Theorem (IV)--(V). 

\item[$(c)$] HFPE (\eqref{HFPE} or equivalently \eqref{HFPE-standard}) obtained from the master equation \eqref{DME} expressed in deformed homotopic space $x_{\gamma,\lambda}$ is a special case of the van Kampen diffusion equation \eqref{van-Kampen} for suitable choices of the mobility and the temperature that now result parameterized homotopically \eqref{conditions}.

\item[$(c)$] Solution of the HFPE \eqref{free-HFPE-2} allows to recover Tsallis and Kaniadakis deformed Fokker-Planck equations studied separately in \cite{daCosta-PRE-2020,GomezPRE-2023} and \cite{CNSNS-2023} along with a mixture behavior measured for the values of the homotopic parameter $0<\lambda<1$, as observed from Figs. 3 and 4. 

\item[$(d)$] Homotopic MSD is symmetrical in the deformation parameter $\gamma$ and for all the cases a superdiffusion is observed, as reported numerically in Fig. 5. This implies that the homotopy tends to increase the rate of the MSD. 

\item[$(e)$] Homotopic SED ratio of Fig. 6, obtained from generalized H-Theorem entropic form \eqref{homotopic-entropy}, manifests a inhomogeneous screening, due to the homotopic PDM \eqref{eq:mlg}. Divergence of the SED at $x=-1/\gamma$ for the Tsallis case is consistent with the localization of the particle (solid gray curve in Fig. 2). 

\end{itemize}

The results of this work are potentially extended to other contexts, for instance the Abe deformation \cite{Abe-2003} (inherited by Abe statistics) and inhomogeneous quantum walks \cite{Linden-2009}.

\section*{Acknowledgments}
Ignacio S. Gomez, Daniel Rocha de Jesus and Ronaldo Thibes acknowledge support from the Department of Exact
and Natural Sciences of the State University of Southwest
Bahia (UESB), Itapetinga, Bahia, Brazil. Ignacio S. Gomez acknowledges support received from the Conselho Nacional de Desenvolvimento Científico e Tecnológico (CNPq), Grant Number
316131/2023-7. Daniel Rocha de Jesus acknowledges support received from CNPq as a scientific initiation scholarship.


\begin{thebibliography}{99}


\bibitem{Hughes-1995}
B. D. Hughes,
{\it Random Walks and Random Environments - Vol I: Random Walks}
(Oxford University Press, New York, 1995).

\bibitem{Rudnick-2004}
J. Rudnick and G. Gaspari,
{\it Elements of the Random Walk - An Introduction for Advanced Students and Researchers}
(Cambridge University Press, New York, 2004).

\bibitem{Tsallis-1988}
C. Tsallis,
J. Stat. Phys. {\bf 52}, 479 (1988).

\bibitem{Tsallis-Bukmann-1996}
C. Tsallis and D. J. Bukman,
Phys. Rev. E {\bf 54}, R2197 (1996).

\bibitem{Baldovin-Robledo-2002}
F. Baldovin and A. Robledo,
Phys. Rev. E {\bf 66}, 045104(R) (2002).

 \bibitem{Kaniadakis-2002}
G. Kaniadakis,
Phys. Rev. E {\bf 66}, 056125 (2002).

\bibitem{Tsallis-2009}
C. Tsallis,
{\it Introduction to Nonextensive Statistical Mechanics}
(Springer, New York, 2009).


\bibitem{Tirnakli-Borges-2016}
U. Tirnakli and E. Borges,
Sci. Rep. {\bf 6}, 23644 (2016).


\bibitem{Nivanen-Mehaute-Wang-2003}
L. Nivanen, A. Le M\'ehaut\'e, and Q. A. Wang,
Rep. Math. Phys {\bf 52}, 437 (2003).

\bibitem{Borges-2004}
E. P. Borges,
Physica A {\bf 340}, 95 (2004).


\bibitem{daCosta-PRE-2020}
da Costa B. G., I. S. Gomez and E. P. Borges,
Phys. Rev. E \textbf{102}, 062105 (2020).

\bibitem{Morse-EPL}
R. N. Costa Filho, G. Alencar, B.S. Skagerstam and J. S. Andrade Jr., EPL \textbf{101}, 10009 (2013).

\bibitem{PDM1} G. Bastard, J. K. Furdyna, and J. Mycielsky, Phys. Rev. B \textbf{12}, 4356 (1975).

\bibitem{PDM2} O. von Roos, Phys. Rev. B \textbf{27}, 7547 (1983).

\bibitem{PDM-DFT} K. Bencheikh, K. Berkane, and S. Bouizane, J. Phys. A: Math.
 Gen. \textbf{37}, 10719 (2004).
 
\bibitem{PDM-SUSY} M. V. Ioffe, E. V. Kolevatova, and D. N. Nishnianidze, Phys.
 Lett. A \textbf{380}, 3349 (2016).
 
\bibitem{PDM-NP} M. Alimohammadi, H. Hassanabadi, and S. Zare, Nucl. Phys.
 A \textbf{960}, 78 (2017).

\bibitem{PDM-NO} K. Li, K. Guo, X. Jiang, and M. Hu, Optik \textbf{132}, 375 (2017).

\bibitem{PDM-LQ} Z. Algadhi and O. Mustafa, Ann. Phys. \textbf{418}, 168185 (2020).

\bibitem{Maike-2021}
M. A. F. dos Santos, I. S. Gomez, B. G. da Costa and O. Mustafa,
Eur. Phys. J. Plus {\bf 136}, 96 (2021).

\bibitem{Gomez-2021}
I. S. Gomez and E. P. Borges,
Lett. Math. Phys. \textbf{111}, 43 (2021).


\bibitem{Tempesta-2011}
P. Tempesta,
Phys. Rev. E {\bf 84},  021121 (2011).

\bibitem{GomezPRE-2023}
I. S. Gomez,  
Phys. Rev. E {\bf 107},  034113 (2023).

\bibitem{Wu-2012}
F. Wu, W. Shi and F. Liu,
\emph{Commun. Nonlinear Sci. Numer. Simul.} \textbf{17}, 2776-2790 (2012).

\bibitem{Gomez-Aguilar-2016}
J. F. Gómez-Aguilar, M. Miranda-Hernández, M. G. López-López, V. M. Alvarado-Martínez, and D. Baleanu,
\emph{Commun. Nonlinear Sci. Numer. Simul.} \textbf{30}, 115-127 (2016).

\bibitem{Pinto-2017}
L. Pinto and E. Sousa,
\emph{Commun. Nonlinear Sci. Numer. Simul.} \textbf{17}, 211-228 (2017).

\bibitem{Maike-2018}
M. A. F. dos Santos and I. S. Gomez,
J. Stat. Mech. {\bf 12}, 123205 (2018).

\bibitem{Renio-2017}
R. S. Mendes, E. K. Lenzi, L. C. Malacarne, S. Picoli, and 
M. Jauregui,
Entropy {\bf 19}(4), 155 (2017).

\bibitem{Jose-2022}
S. Jose, D. Mandal, M. Barma, and K. Ramola,
Phys. Rev. E {\bf 105}, 064103 (2022).


\bibitem{CNSNS-2023}
I. S. Gomez, B. G. da Costa and M. A. F. Santos,
Commun. Nonlin. Sci. Num. Sim. \textbf{119}, 107131 (2023).

\bibitem{Oppenheim-1977}
I. Oppenheim, K. E. Schuler and G. H. Weiss,
{\it Stochastic Processes in Chemical Physics
The Master Equation}
(ISBN: 9780262150170, 1977).

\bibitem{Risken-1989}
H. Risken,
{\it The Fokker-Planck Equation: Methods of Solution and Applications}
(Springer, Berlin, 1989).

\bibitem{vanKampen-1987}
N. G. van Kampen,
Z. Phys. B Condensed Matter {\bf 68}, 135 (1987).


\bibitem{Abe-2003}
S. Abe, 
\emph{J. Phys. A: Math. Gen.} \textbf{36} 8733 (2003).

\bibitem{Linden-2009}
N. Linden and J. Sharam,
\emph{Phys. Rev. A} {\bf 80}, 052327 (2009).







\end{thebibliography}
\end{document}